\documentclass[aps,prd,twocolumn,superscriptaddress,longbibliography,nofootinbib]{revtex4-2}
\usepackage[utf8]{inputenc}
\usepackage{color}
\usepackage{graphicx}
\usepackage{subfigure}
\usepackage{gensymb}
\usepackage{xcolor}
\usepackage[normalem]{ulem}
\usepackage[pdftex,breaklinks,colorlinks,linkcolor=blue,citecolor=teal,anchorcolor=red,urlcolor=cyan]{hyperref}
\usepackage{multirow}

\usepackage{amsmath,amssymb,amsfonts}
\DeclareMathOperator\arctanh{arctanh}

\def\prl{Phys. Rev. Lett.}
\def\prd{Phys. Rev. D}

\def\apj{Astrophys. J.}

\def\apjl{Astrophys. J. Lett.}

\def\pr{Phys. Rev.}

\def\pau_p{Prog. Theor. Phys.}

\def\mnras{Mon. Not. R. Astron. Soc.}

\def\nat{Nature}
\def\sovast{Soviet Ast.}
\def\jcap{J. Cosmology Astropart. Phys}

\def\l{\ell}
\def\rhc{{\rho}_c}
\def\drhot{\rho_c^{(2)}}
\def\drhof{\rho_c^{(4)}}
\def\Pc{P_c}
\def\P2{P_c^{(2)}}
\begin{document}

\title{Primordial black holes, gravitational wave beats, and the nuclear equation of state}

\author{Thomas W.~Baumgarte}
\email{tbaumgar@bowdoin.edu}
\affiliation{Department of Physics and Astronomy, Bowdoin College, Brunswick, Maine 04011, USA}

\author{Stuart L.~Shapiro}
\email{slshapir@illinois.edu}
\affiliation{Department of Physics, University of Illinois at Urbana-Champaign, Urbana, Illinois 61801}
\affiliation{Department of Astronomy and NCSA, University of Illinois at Urbana-Champaign, Urbana, Illinois 61801}

\begin{abstract}
Primordial black holes (PBHs), if trapped in neutron stars (NSs), emit a characteristic continuous, quasiperiodic gravitational wave (GW) signal as they orbit inside the host star.  We identify a specific and qualitatively new feature of these signals, namely quasiperiodic beats caused by the precession of noncircular PBH orbits.  We demonstrate numerically and analytically that the beat frequency depends rather sensitively on the NS structure, so that hypothetical future observations with next-generation GW detectors could provide valuable constraints on the nuclear equation of state.
\end{abstract}

\maketitle

{\em Primordial black holes} (PBHs), first proposed by \cite{ZelN67,Haw71}, may have formed in the early Universe, and may contribute to or even make up most of its dark-matter content (see also \cite{CarH74}).  While observational constraints on PBHs limit their possible contribution to the dark matter in some mass ranges, they remain viable candidates in other mass windows, including between about $10^{-16} M_\odot$ and $10^{-10} M_\odot$ as well as around $10^{-6} M_\odot$ (see, e.g., \cite{Khl10,CarK20,CarKSY21} for reviews and details).  

If PBHs exist, some of them are likely to interact with stars and other celestial objects.  Such interactions have been invoked as possible origins of several astrophysical phenomena, including the 1908 Tunguska event in Siberia \cite{JakR73} (but see \cite{BeaT74}), neutron star (NS) implosions and ``quiet supernovae" \cite{FulKT17,BraL18}, fast radio bursts \cite{FulKT17,AbrBW18,AmaS23}, the formation of low-mass stellar black holes \cite{Tak18,TakFK21,AbrBUW22,OncMGG22}, microquasars \cite{Tak19}, and the origin of supermassive black holes (e.g.~\cite{BamSDFV09}), possibly via the formation of PBH clusters \cite{BelDEGGKKRS14,BelDEEKKKNRS19}.  Gravitational-wave (GW) signatures of PBHs have been surveyed recently in \cite{Bagetal23}, and the prospect of detecting PBHs using solar-system ephemerides has been discussed in \cite{BerCDVC23,TraGLK23} and references therein.

A collision with a star results in the PBH being gravitationally bound if it loses a sufficient amount of energy in the encounter, which is most likely to happen in collisions with NSs (see, e.g., \cite{CapPT13,AbrBW18,MonCFVSH19,GenST20}).  The PBH may still emerge from the star, but can no longer escape to infinity.  Losing more energy in subsequent passages, the PBH at some point remains completely inside the star, settles down toward its center, accretes stellar material, and ultimately induces the dynamical collapse of the host star (see \cite{EasL19,RicBS21b,SchBS21} for numerical simulations).  While the expected event rates are small (see, e.g., \cite{Abretal09,CapPT13,MonCFVSH19,HorR19,ZouH22} as well as  Section I in the supplementary material for estimates) they depend strongly on a number of assumptions and may be more favorable in special environments, e.g.~globular clusters and galactic centers.  Small black holes may also form inside neutron stars from the  collapse of other dark-matter particles (e.g.~\cite{GolN89,McDYZ12,BraL18,EasL19}), or be captured by neutron stars by other processes (e.g.~\cite{BamSDFV09,PanL14,HorR19,GenST20}).

While the PBH spirals toward the center of the NS it emits gravitational radiation that -- at least in principle -- may be observable by next-generation GW detectors, and that would reveal information about the stellar structure \cite{HorR19}.  The authors of \cite{ZouH22,GaoDGZZZ23}, for example, examined this scenario assuming circular orbits.  Since the PBH typically enters the host star on a noncircular orbit, and since the retarding forces inside the star may not circularize the orbit (see, e.g., \cite{Hof85,SzoML22}), the PBH's orbit is likely to remain eccentric (see also \cite{BauS24c} for a numerical demonstration).  In this letter we discuss a qualitatively new feature of such noncircular orbits, namely continuous, {\em quasiperiodic GW beats}. These beats are caused by a precession of the PBH's orbit inside the star, the GW frequency for which is superimposed on the higher frequency arising from a single orbit. The resulting GW envelopes for the two GW polarizations are exactly out of phase, so that the GW signal alternates between being dominated by one or the other polarization. In the stellar interior, both Newtonian and relativistic effects contribute to this precession, but we find that the latter dominate in NSs.  As we demonstrate both numerically and analytically, the rate of the precession, and hence the beat frequency, depends rather strongly on the NS structure, so that {\em a future observation of such a GW beat could provide strong constraints on the nuclear equation of state (EOS), let alone confirmation of the capture by a NS of a smaller and lower-mass intruder.} 

\begin{figure*}
    \centering
    \includegraphics[width=0.9 \textwidth]{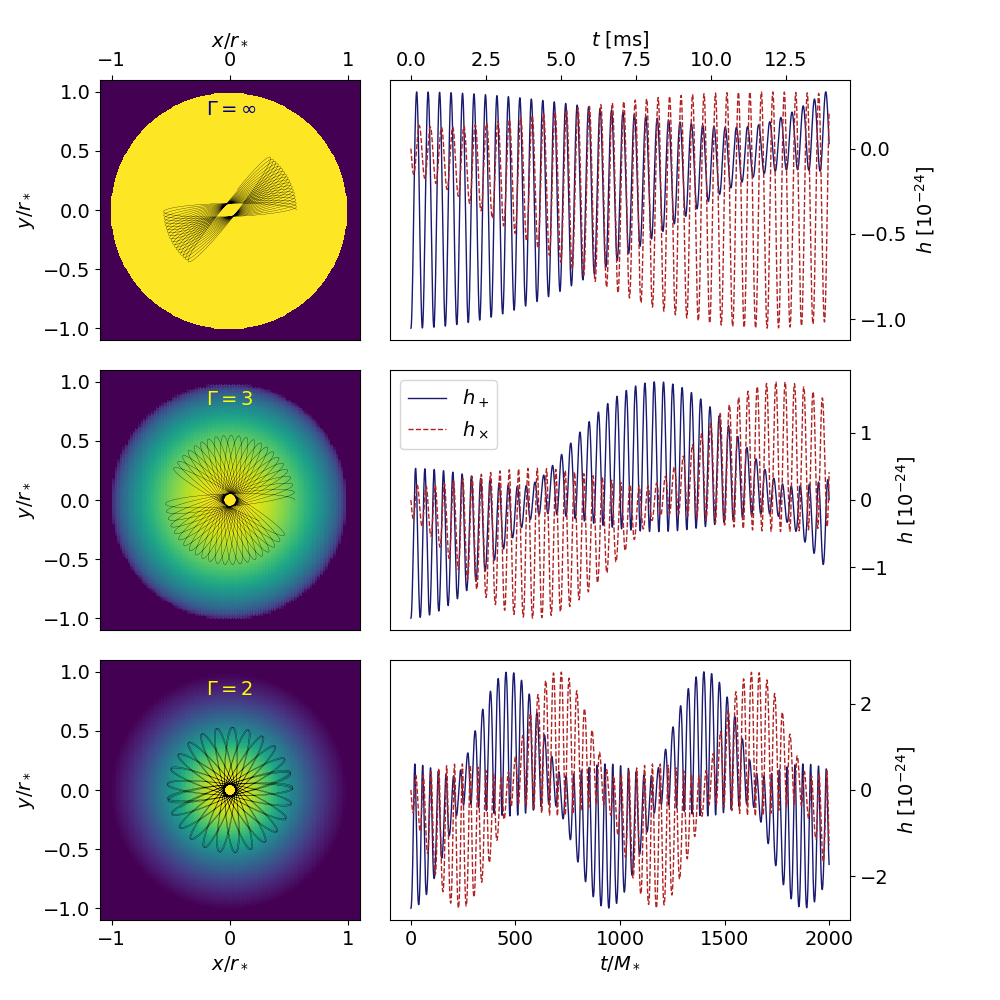}
    \caption{Numerical examples of orbits inside NSs governed by different EOSs (left panels) together with the emitted GW signal as a function of time (right panels).  The color shading in the left panels represents the density distribution inside the star. The top row shows, as the extreme limit, a constant-density star, corresponding to $\Gamma = \infty$.   The middle and bottom rows show results for $\Gamma = 3$ and $\Gamma = 2$ polytropes. The GW amplitudes $h_+$ and $h_\times$ in the right panels are scaled for a star of mass $M_* = 1.4 M_\odot$ and hosting a black hole of mass $m = 10^{-6} M_*$ at a distance of $d = 10$ kpc. The time $t$ in the right panel denotes time as measured by a distant observer, and is provided both in units of the the stellar mass $M_*$ (bottom axis) and in terms of ms (upper axis). (See \cite{animation} for an animation.)}
    \label{fig:orbits}
\end{figure*}

As a numerical demonstration we show in Fig.~\ref{fig:orbits} characteristic orbits of PBHs inside NSs governed by three different EOSs varying in stiffness, together with their associated GW signals.  We adopt a simple particle test-mass approximation to describe the PBH moving on geodesics in the gravitational field of a relativistic star. As dynamical friction and accretion drag forces are small perturbations that operate on secular timescales much longer than orbital times (e.g.~\cite{Abretal09,GenST20}), we can probe the precession by neglecting these forces and examining a few orbits via geodesics.  We assume the stars to be governed by a  polytropic EOS
\begin{equation} \label{polytrope}
P = K \rho_0^\Gamma.
\end{equation}
Here $P$ is the pressure, $\rho_0$ the rest-mass density, $K$ a constant, and the adiabatic exponent $\Gamma = 1 + 1/n$ may be expressed in terms of the polytropic index $n$.  We construct the stellar models by solving the Oppenheimer-Volkoff (OV) equations, adjusting the central density so that the stellar compaction is always given by $G M_*/(c^2 R_*) = 1/6$, where $M_*$ is the star's total gravitational mass and $R_*$ is areal radius (see Section II in the supplementary material for details).   In Fig.~\ref{fig:orbits} we show examples for a star with constant total mass-energy density $\rho$ (corresponding to $\Gamma = \infty$), as well as $\Gamma = 3$ and $\Gamma = 2$ polytropes, which serve as examples of both highly and moderately stiff candidates for the NS EOS.

We then solve the relativistic geodesic equations in order to track the PBH's orbit.  As discussed in Section III of the supplementary material, we choose to solve these equations in terms of the isotropic metric and radius $r$ rather than the interior Schwarzschild metric and areal radius $R$, although the conversion is straightforward.  We always start orbits with vanishing radial speed $u^r = 0$ at an initial (areal) radius $R(0) = R_{\rm frac} R_*$ and with angular momentum $\l = \l_{\rm frac} \l_{\rm circ}$, where $\l_{\rm circ}$ is the angular momentum corresponding to a circular orbit at radius $R(0)$.  For the examples shown in Fig.~\ref{fig:orbits} we used $R_{\rm frac} = 0.6$ and $\l_{\rm frac} = 0.1$.  All orbits are confined to a plane, which we arbitrarily take to be the $x-y$ plane.  We also evaluate the leading-order GW signals $h_+$ and $h_\times$ along the $z-$axis using the quadrupole formalism \cite{PetM63,Pet64,MisTW73}.

As can be seen in the left column of Fig.~\ref{fig:orbits}, the rate at which the PBH's orbit precesses depends strongly on the structure of the host star, and hence its EOS.  For all three examples we show the orbits for a time span $\Delta t = 2000 M_*$, which corresponds to about 14 ms for $M_* = 1.4 M_\odot$. All orbits start out on the positive $x$-axis.  For the constant-density star in the top row, the orbit has rotated by just over $45\degree$ during this time, while for $\Gamma = 3$ it has rotated by a little less than $180\degree$, and a little over $360\degree$ for $\Gamma = 2$.

If the GW signal is dominated by one polarization initially (in our case $h_+$), then it will be dominated by the other polarization after the orbit has rotated by $45\degree$, and will return to the original polarization after a rotation through $90\degree$.  The resulting GW beats, and their dependence on the NS structure, can be seen in the right panels of Fig.~\ref{fig:orbits}.  Specifically, we observe that the signal has shifted from being dominated by $h_+$ to being dominated by $h_\times$ for the constant density star, while for $\Gamma = 3$ it has gone from $h_+$ to $h_\times$ and then back to $h_+$ almost twice, and just over four times for $\Gamma = 2$.

The above behavior can be understood in part in the context of {\em Bertrand's theorem} (\cite{Ber73}, see also \cite{Gol80}), which states that for potentials $V(R) = V_0 + k R^m$, where $V_0$ and $k$ are constants, only the exponents $m = -1$ and $m = 2$ will always result in closed orbits.  

The former exponent, $m = -1$, corresponds to a Newtonian point-mass potential, which is the leading order term in the potential in the exterior of a star.  In general, deviations from this exterior potential result both from relativistic corrections as well as several Newtonian effects, including tidal and rotational deformations of the star (which are not present for our static spherical stars).  These deviations result in precession of the orbit, including the well-known relativistic perihelion advance of Mercury (see Section IV.A of the supplemental material).

The latter exponent, a harmonic-oscillator potential with $m = 2$, is realized in the interior of a Newtonian constant-density star, for which $M(R) = 4 \pi \rho R^3 / 3$ and hence $V(R) = V_0 + 2 \pi G \rho R^2 / 3$. In this case, deviations result from both density nonuniformities and relativistic corrections.  While we do not expect any precession, either in the interior or exterior, for a  homogeneous spherical star in Newtonian gravitation, relativistic effects will cause precession for such a star both in the exterior and interior.

Even for an inhomogeneous star, the stellar core becomes increasingly homogeneous as $R \rightarrow 0$.  In Newtonian gravitation, the PBH's orbit therefore starts with a nearly closed orbit with $m = -1$ far outside the NS, and ends with a tighter, nearly closed orbit with $m = 2$ well inside the NS -- quite remarkably realizing both cases of Bertrand's theorem as extreme limits.  Moreover, since the approximately homogeneous core extends to larger radii for stiffer EOSs than for softer EOSs, we expect that, for a given orbital radius, the precession will be slower for a stiffer EOS than for a softer EOS in Newtonian theory.  This precession rate variation is also found in general relativity, as revealed in the numerical examples of Fig.~\ref{fig:orbits}. 

We can gain analytical insight into the above effects by considering small perturbations of circular orbits.  Geodesics in static and spherically symmetric spacetimes possess two conserved constants of motion, namely the energy per unit mass $e = - u_t$ and the angular momentum per unit mass $\l = u_{\varphi}$.  Using the normalization of the four-velocity $u^a$, $g_{ab} u^a u^b = - 1$, a first integral of the equations of motion can be written in the form
\begin{equation} \label{first_int}
\frac{1}{2} (u^R)^2 = E - V_{\rm eff}(R),
\end{equation}
where the constant $E \equiv (e^2-1)/2$ plays the role of the kinetic energy at infinity for velocities $v_{\rm \infty}  \ll 1$, and where we split the effective potential $V_{\rm eff}(R)$ into the two terms
\begin{equation}
V_{\rm eff}(R) = \frac{\l^2}{2 R^2} + V(R).
\end{equation}
In many cases (e.g., a Newtonian point-mass) $V(R)$ is independent of $e$ and $l$, but we now allow this term to depend on these two constants, which is the case for the orbits in general relativity considered here. In the following we assume the orbit to be in the equatorial plane, so that $\theta = \pi/2$, and we provide details of how $V(R)$ can be determined in Section IV of the Supplement.

For a stable circular orbit at (areal) radius $R_0$ the effective potential $V_{\rm eff}(R)$ must take a minimum there, so that we have
$V'(R_0) \equiv (dV/dR)_{R_0} = \l^2/R_0^3$.  Since $\l = g_{\varphi\varphi} u^\varphi = R^2 d\varphi / d\tau$, the proper time $\tau_{\varphi}$ needed to complete one orbit, i.e.~to advance from an angle $\varphi_0$ to $\varphi_0 + 2 \pi$, is given by
\begin{equation}
\tau_\varphi = \frac{2 \pi R_0^2}{\l} = 2 \pi \left( \frac{R_0}{V'(R_0)} \right)^{1/2}.
\end{equation}
In the vicinity of a stable circular orbit we may approximate the effective potential as a parabola $V_{\rm eff}(R) \simeq V_{\rm eff}(R_0) + k \eta^2/2$, where $k \equiv V_{\rm eff}''(R_0) > 0$ and $\eta \equiv R - R_0$.  Inserting these into (\ref{first_int}) and taking a derivative with respect to proper time $\tau$ results in the harmonic-oscillator equation $\ddot \eta + k \eta = 0$ for $\eta$, where the double dot denotes a second derivative with respect to $\tau$.  Accordingly, the proper time $\tau_R$ needed to travel from the orbit's pericenter to the apcenter and back to the pericenter is given by 
\begin{equation}
\tau_R = \frac{2 \pi}{k^{1/2}} =  
2 \pi \left( \frac{R_0}{3 V'(R_0) + R_0 V''(R_0)} \right)^{1/2},
\end{equation}
and the ratio between the two times $\tau_R$ and $\tau_\varphi$ is
\begin{equation} \label{tau_ratio}
\frac{\tau_R}{\tau_\varphi} = \left( \frac{V'(R_0)}{3 V'(R_0) + R_0 V''(R_0)} \right)^{1/2}.
\end{equation}
For a Newtonian point-mass potential 
$V(R) = p R^{-1}$, where $p$ is a constant, we have $\tau_R = \tau_\varphi$, as expected for Kepler orbits.  For a harmonic-oscillator potential $V(R) = p R^2$ we have $\tau_R = \tau_\varphi / 2$, so that the orbit, which is centered at and symmetric about the origin, features two pericenters in each revolution.  According to Bertrand's theorem, these two cases are the {\em only} potentials that lead to closed orbits.  

While perturbations of the point-mass potential in the stellar exterior are familiar -- yielding, for example, the relativistic perihelion advance of Mercury -- we now focus on the stellar interior.  Specifically, we show in Section IV.B of the Supplement that, in the vicinity of the center, the potential $V(R)$ can be written in the form 
\begin{equation} \label{pot}
V(R) = V_0 + p R^2 + q R^4 + {\mathcal O}(R^6),
\end{equation}
where $V_0$, $p$, and $q$ are constants.  Inserting (\ref{pot}) into (\ref{tau_ratio}) and assuming $q R_0^2 \ll p$ we find, to leading order,
\begin{equation}
\frac{\tau_{R}}{\tau_\varphi} \simeq \frac{1}{2} \left(1 - \frac{q R_0^2}{2 p} \right).
\end{equation}
As the black hole advances from one pericenter to the next, its positional angle $\varphi$ therefore advances by
\begin{equation} \label{advance1}
\Delta \varphi = \frac{d \varphi}{d\tau} \tau_R - \pi \simeq - \frac{\pi}{2} \frac{q R_0^2}{p} 
\end{equation}
beyond the angle $\pi$ that would result in a closed orbit. 

We refer the reader to Section IV of the Supplement where the constants $p$ and $q$ are evaluated in general relativity for nearly circular orbits.  Here we present a more transparent Newtonian treatment in order to illustrate the key ingredients.  In the vicinity of the stellar center we may approximate the density as $\rho(R) \simeq \rho_c + \rho^{(2)} R^2 /2$, where $\rho^{(n})_c \equiv (d^n \rho/dR^n)_{R = 0}$. Integrating once we find the enclosed mass $M(R)$, and integrating again we obtain the potential
\begin{equation}
V_{\rm Newton}(R) \simeq V_0 + \frac{2 \pi G}{3} \rho_c R^2 + \frac{\pi G}{10} \rho_c^{(2)} R^4.
\end{equation}
Comparing with (\ref{pot}) we identify both $p$ and $q$ and compute
\begin{equation} \label{advance2}
\Delta \varphi_{\rm Newton} \simeq - \frac{3 \pi}{40} \frac{\rho^{(2)}_c R_0^2}{\rho_c}.
\end{equation}
Evidently, the Newtonian pericenter advance is related to the degree of inhomogeneity, consistent with our discussion above.

\begin{table*}[]
    \centering
    \begin{tabular}{c|c||c|c|c||c|c|c|c|c|c|c|c}
        \multicolumn{2}{c||}{} & \multicolumn{3}{|c||}{Newton} & \multicolumn{8}{|c}{GR} \\
        \cline{1-13} 
         $\Gamma$ & $n$ & $\delta$ & $\Delta \varphi / (R_0 / R_*)^2$ & $\Delta \varphi$ & $\delta$ & $\Delta \varphi_{\rm num}$ &
        $\Delta \varphi_{\rm ana}$ & ${\mathcal N}_{\rm GW}$ & $\tau_{\varphi}$~[ms] & $\tau_{\rm prec}$~[ms]  & $f_{\rm orb}^{\rm GW}$~[kHz]  & $f_{\rm beat}^{\rm GW}$~[Hz] \\ \hline
         1.75 & 1.33 & 4.89 & 1.26 & 0.00315 & 7.46 & 0.0161 & 0.0162 & 194 & 0.194 & 18.9 & 5.56 & 28.7 \\ 
         2.0 & 1.0 & 3.29 & 0.775 & 0.00193 & 3.98 & 0.00841 & 0.00844 & 372 & 0.273 & 51.0 & 4.48 & 12.1 \\
         2.25 & 0.8 & 2.60 & 0.556 & 0.00139 & 2.94 & 0.00608 & 0.00612 & 513 & 0.321 & 82.4 & 4.01 & 7.81 \\
         2.5 & 0.67 & 2.23 & 0.432 & 0.00108 & 2.43 & 0.00494 & 0.00498 & 630 & 0.354 & 111 & 3.73 & 5.91 \\
         2.75 & 0.57 & 2.00 & 0.352 & 0.00088 & 2.14 & 0.00427 & 0.00430 & 731 & 0.379 & 138 & 3.54 & 4.85 \\
         3.0 & 0.5  & 1.84 & 0.298 & 0.00075 & 1.94 & 0.00382 & 0.00385 & 816 & 0.398 & 163 & 3.41 & 4.18 \\ 
         3.25 & 0.44 & 1.72 & 0.257 & 0.00064 & 1.80 & 0.00351 & 0.00353 & 890 & 0.414 & 184 & 3.31 & 3.72 \\ 
         $\infty$ & 0 & 1 & 0 & 0 & 1 & 0.00164 & 0.00165 & 1904 & 0.553 & 527 &  2.62 & 1.38  
    \end{tabular}
    \caption{Numerical and analytical data for polytropic stellar models and pericenter advances for nearly circular orbits close to the center of a stellar host with compaction $G M_*/(c^2 R_*) = 1/6$.  We list, for different values of $\Gamma = 1 + 1 / n$, Newtonian values of the central condensation $\delta = \rho_c / \bar \rho$ (which depends on the $\Gamma$ alone) and the Newtonian estimate $\Delta \varphi / (R_0/R_*)^2$, adopting numerical solutions to the Lane-Emden equation, together with $\Delta \varphi$ for $R_0 / R_* = 0.05$.  For the relativistic data we computed $\delta$ from solutions to the OV equations.  Adopting $R_0 / R_* = 0.05$ again we computed $\Delta \varphi_{\rm num}$ from numerical solutions to the geodesic equations, using $\ell_{\rm frac} = 0.99$ for nearly circular orbits, and $\Delta \varphi_{\rm ana}$ analytically as presented in the Supplement.  We also list the number of GW wave cycles ${\mathcal N}_{\rm GW}$ completed during a beat cycle (see Eq.~\ref{cycles}), and, assuming a host star with mass $M_* = 1.4 M_\odot$, the orbital time $\tau_\varphi$ as well as the precession time $\tau_{\rm prec}$.  In the last two columns we provide the corresponding GW frequencies $f_{\rm orb}^{\rm GW} \simeq 2 (u^t \tau_{\varphi})^{-1}$ and $f_{\rm beat}^{\rm GW} \simeq (u^t \tau_{\rm prec})^{-1}$ as measured by a distant observer.}
    \label{tab:polytropes}
\end{table*}

We may evaluate the term $\rho^{(2)}_c$ in (\ref{advance2}) using the Newtonian equations of hydrostatic equilibrium.  For a polytropic EOS (\ref{polytrope}) we find
\begin{equation} \label{denspp}
    \rho^{(2)}_c = - \frac{4 \pi}{3} \frac{G \rho_c^2}{a_c^2},
\end{equation}
where $a = (\Gamma P / \rho)^{1/2}$ is the (Newtonian) speed of sound.  Finally we may use the central condensation $\delta \equiv \rho_c/\bar \rho$, where $\bar \rho = 3 M_*/(4 \pi R_*^3)$ is the average density, to rewrite the Newtonian pericenter advance as
\begin{equation} \label{advance3}
\Delta \varphi_{\rm Newton} \simeq \frac{3 \pi}{40} \frac{G M_*}{a_c^2 R_*} \left(\frac{R_0}{R_*} \right)^2 \delta.
\end{equation}
For Newtonian polytropes $\delta$ depends on $\Gamma$ only (see Table \ref{tab:polytropes} for specific values).  For smaller $\Gamma$, $\delta$ is larger and $a_c^2$ smaller (for a given compaction $GM_*/(c^2 R_*)$); we therefore  see that the pericenter advance is larger for a softer EOSs (for a given value of $R_0/R_*$).

While the above analysis captures the leading-order Newtonian terms, we have found that the pericenter advance in the NSs considered here is dominated by relativistic terms.   However, the pericenter advance's dependence on the EOS's stiffness is similar to that observed from the above Newtonian analysis even in the context of general relativity -- namely, a softer EOS will lead to a more rapid precession of the orbit, and therefore to higher-frequency GW beats.  This can be observed in Fig.~\ref{fig:orbits} as well as in the Table \ref{tab:polytropes}, where we list pericenter adnvances $\Delta \varphi$ for nearly-circular orbits ($\l_{\rm frac} = 0.99)$ close to the center ($R_{\rm frac} = 0.05$) for a range of polytropic indices.  We compare numerical results from the integration of the geodesic equation with analytical results from the perturbation of nearly-circular orbits and find excellent agreement.

The range of polytropic exponents $\Gamma$ listed in Table \ref{tab:polytropes} roughly covers values adopted in piecewise-polytropic approximations for candidate nuclear EOSs (see Table III in \cite{ReaLOF09}; note in particular the larger range of values for $\Gamma_3$, which governs the high-density core). The resulting values of $\Delta \varphi$ show significant variation, suggesting that a potential observation of the resulting GW beats would provide a sensitive probe of the EOS.  

From the pericenter advance $\Delta \phi$ we may also compute the precession frequency.  Since we defined $\Delta \varphi$ as the (excess) advance from one pericenter to the next, and since, to leading order, orbits in the stellar interior feature two pericenters per orbit, the angular precession frequency measured locally is given by
$\Omega_{\rm prec} = 2 \Delta \varphi / \tau_{\varphi}$,
where $\tau_\varphi$ is the orbital (proper) period. Related to the precession frequency is the (proper) precession period $\tau_{\rm prec}$ that it takes either of the two GW polarization amplitudes to go through one complete cycle, i.e.~for the pericenter to advance by an angle $\pi$,\footnote{In the bottom panel of Fig.~\ref{fig:orbits}, for example, $\tau_{\rm prec} \simeq 1000 M_*$.}
\begin{equation} \label{tau_prec}
\tau_{\rm prec} = \frac{\pi}{\Omega_{\rm prec}} = \frac{\pi}{2 \Delta \varphi} \tau_\varphi.
\end{equation}
The number of orbits completed during a precession period $\tau_{\rm prec}$ is therefore ${\mathcal N}_{\rm orbit} = \tau_{\rm prec}/\tau_{\rm \varphi} = \pi/(2 \Delta \varphi).$  Since, during one revolution, the GW signal completes two cycles, the number of such GW cycles completed as either GW polarization goes through a full beat cycle associated with the GW envelope is given by 
\begin{equation} \label{cycles}
{\mathcal N}_{\rm GW} = 2 {\mathcal N}_{\rm orbit} = \frac{\pi}{\Delta \varphi}.
\end{equation}
For the orbits in Fig.~\ref{fig:orbits}, for example, we found $\Delta \phi = 0.0255$, 0.0597, and 0.127 for $\Gamma = \infty$, 3, and 2, respectively, resulting in ${\mathcal N}_{\rm GW} = 123$, 52.6, and 24.8.  In Table \ref{tab:polytropes} we also provide data for ${\mathcal N}_{\rm GW}$, $\tau_\varphi$, and $\tau_{\rm prec}$ for the nearly circular orbits considered there.

The above periods are proper times as measured by an observer co-moving with the PBH.  For the near-circular orbits in Table \ref{tab:polytropes} we may simply multiply these periods with $u^t$ in order to obtain the corresponding coordinate time periods.  In Table \ref{tab:polytropes} we list the resulting GW frequency  associated with a single orbit, $f^{\rm GW}_{\rm orb} \simeq 2 /(u^t \tau_\varphi)$, and the GW beat frequency $f^{\rm GW}_{\rm beat} \simeq 1/(u^t \tau_{\rm prec})$, both as measured by a distant observer.

To summarize, we discuss quasiperiodic GW beats as a qualitatively new feature of continuous GW signals emitted by PBHs captured inside NSs.  The beats are due to orbital precession, which is caused both by relativistic effects and density nonuniformity.  Adopting a polytropic EOS we demonstrate both numerically and analytically that the beat frequency depends quite strongly on the structure of the NS and hence the stiffness of the EOS.  For the NSs considered here {\em the precession rate and beat frequency are largely due to relativistic gravitation}, so that a Newtonian treatment would significantly underestimate the effect.  If such beats were to be observed by next-generation GW detectors, e.g.~the Einstein Telescope \cite{et}, the Cosmic Explorer \cite{ce}, or the Neutron Star Extreme Matter Observatory (NEMO) \cite{NEMO}, they would therefore provide valuable constraints on the nuclear EOS.

Clearly, the beat frequency also depends on the radius and eccentricity of the PBH's orbit, which would have to be found independently.  The latter is related to the relative maximum and minimum amplitudes in each one of the GW polarizations, and it may be possible to determine the radius from the prior inspiral signal.  Knowing these orbital parameters, as well as the host star's compaction, an observed beat frequency could then be compared with those found for orbits inside general relativistic stellar models constructed for candidate nuclear EOSs.  While the GW amplitude depends on the PBH mass, the precession frequency does not.

\acknowledgements

We would like to thank Steve Naculich, David Kaiser, and Keith Riles for helpful conversations. T.W.B.~gratefully acknowledges the hospitality of the University of Arizona's Steward Observatory.  This work was supported in part by National Science Foundation (NSF) grants PHY-2010394 and PHY-2341984 to Bowdoin College, as well as NSF grants PHY-2006066 and PHY-2308242 to the University of Illinois at Urbana-Champaign.


%

\newpage
\begin{center}
{\bf \large Supplemental Material}
\end{center}

In this supplemental material we provide additional information on expected event rates (Section \ref{sec:eventrates}), the construction of the stellar background models (Section \ref{sec:background}), the equations of motion (Section \ref{sec:eom}), as well as the perturbation of nearly circular orbits (Section \ref{sec:perturbations}).   

\section{Event rates}
\label{sec:eventrates}

Rates of collisions between neutron stars and PBHs have been estimated by a number of different authors (e.g.~\cite{Abretal09,CapPT13,MonCFVSH19,HorR19,ZouH22} and references therein).  Here we briefly review some of the key arguments required for a rough estimate.

An individual neutron star will be hit by PBHs at a rate of approximately
\begin{equation} \label{N_dot_1}
\dot {\mathcal N}_{\rm NS} \simeq \sigma v_\infty n_{\rm PBH},
\end{equation}
where $\sigma$ is the cross-section for the collision, $v_\infty$ the relative speed at large distances, and $n_{\rm PBH}$ the number density of the PBHs.  

We estimate the cross-section $\sigma$ from
\begin{equation}
\sigma = \pi R_*^2 \left( 1 + \left( \frac{v_{\rm esc}}{v_\infty} \right)^2 \right)
\end{equation}
(see, e.g., \cite{Abretal09,HorR19}), where 
$
v_{\rm esc} = (2 G M_* / R_*)^{1/2}
$
is the escape speed from the neutron star with mass $M_*$ and radius $R_*$.  For a typical neutron star with $M_* \simeq 1.4 M_\odot$ and $R_* \simeq 6 G M_* / c^2 \simeq 12.5$ km we have $v_{\rm esc} \simeq 0.6 \, c$.  Further adopting $v_\infty \simeq 220$ km/s as a typical speed for halo dark-matter particles (see, e.g., \cite{EvaOM18}) we obtain
\begin{equation}
\sigma \simeq 3 \times 10^{18} \, \mbox{cm}^2.
\end{equation}
We take this value as a constant in the following, ignoring possible changes in $v_\infty$.

The PBH number density $n_{\rm PBH}$ is
\begin{equation}
n_{\rm PBH} \simeq f_{\rm PBH} \frac{\rho_{\rm DM}}{m},
\end{equation}
where $f_{\rm PBH} = \Omega_{\rm PBH}/\Omega_{\rm DM}$ is the mass fraction of the Universe's dark matter content in PBHs, $\rho_{\rm DM}$ is the mass density of dark matter, and we assume that all PBHs have the same mass $m$.  Locally, in the solar neighborhood, the mass density of dark matter is approximately
\begin{equation}
\rho_{\rm DM}^{\rm loc} \simeq 1.3 \times 10^{-2} M_\odot \, \mbox{pc}^{-3} \simeq 10^{-24} \mbox{g}\,\mbox{cm}^{-3}
\end{equation}
(see, e.g., \cite{McKPH15}).  

Inserting the above expressions and numbers into (\ref{N_dot_1}) we obtain
\begin{equation} \label{N_dot_2}
\dot {\mathcal N}_{\rm NS} \simeq 3 \times 10^{-26} \, \mbox{s}^{-1} \, f_{\rm PBH} \left( \frac{10^{-6} M_\odot}{m} \right) \left( \frac{\rho_{\rm DM}}{\rho_{\rm DM}^{\rm loc}} \right).
\end{equation}
The Galactic collision rate is then approximately
\begin{equation} \label{N_dot_gal}
\dot {\mathcal N}_{\rm Gal} \simeq 10^{-9} \, \mbox{yr}^{-1} \,  f_{\rm PBH} \left( \frac{10^{-6} M_{\odot}} {m} \right) \left( \frac{\rho_{\rm DM}^{\rm ave}}{\rho_{\rm DM}^{\rm loc}} \right) \left( \frac{N_{\rm NS}}{10^9} \right),
\end{equation}
where $N_{\rm NS}$ is the number of neutron stars in our Galaxy, and where $\rho_{\rm DM}^{\rm ave}$ represents a Galactic average of the dark-matter mass density.  Near the Galactic center, in particular, the dark matter density is expected to be larger than its local value in the solar neighborhood by several orders of magnitude (see, e.g., \cite{BerM05}).

We note that the above event rates are inversely proportional to the PBH mass $m$, meaning that a smaller value of $m$ will result in a larger Galactic event rate.  However, the strain $h$ of the emitted gravitational wave signal is {\em proportional} to $m$, so that the distance $d_{\rm max}$ to the furthest observable event, for a given detector sensitivity $h_{\rm min}$, is also proportional to $m$.  Accordingly, the observable volume scales with $m^3$, and the total number of observable events -- assuming a uniform distribution of potential sources -- is proportional to $m^2$.

As a specific example, consider the orbits shown in Fig.~1 of the main paper, for which 
\begin{align}
h & \simeq 10^{-24} \left( \frac{m}{10^{-6} M_\odot} \right) \left( \frac{10 \,\mbox{kpc}}{d} \right).
\end{align}
Given a detector sensitivity $h_{\rm min}$ for a continuous gravitational wave signal at a frequency of a few kHz we can estimate $d_{\rm max}$ from requiring that $h \geq h_{\rm min}$, i.e.
\begin{equation}
d_{\rm max} = 10^2 \, \mbox{Mpc} \left( \frac{10^{-28}}{h_{\rm min}} \right) \left( \frac{m}{10^{-6} M_\odot} \right).
\end{equation}
Denoting the number density of galaxies by $n_{\rm gal}$ we can then estimate the total number of observable events from
\begin{widetext}
\begin{align} \label{N_dot_tot}
\dot {\mathcal N}_{\rm tot} \, \simeq \, \frac{4 \pi}{3} d_{\rm max}^3 n_{\rm gal} \dot {\mathcal N}_{\rm Gal} 
\, \simeq \, 4 \times 10^{-5} \, \mbox{yr}^{-1} \, f_{\rm PBH} \left( \frac{10^{-28}}{h_{\rm min}} \right)^3 \left( \frac{m}{10^{-6} M_\odot} \right)^2 \left( \frac{\rho_{\rm DM}^{\rm ave}}{\rho_{\rm DM}^{\rm loc}} \right) \left( \frac{N_{\rm NS}}{10^9} \right) \left( \frac{n_{\rm gal}}{0.01\,\mbox{Mpc}^{-3}} \right),
\end{align}
\end{widetext}
where we have assumed that the Galactic event rates for the Milky Way are typical for all galaxies accounted for in $n_{\rm gal}$.  

While the event rates (\ref{N_dot_tot}) are small, even given our optimistic scaling to $h_{\rm min} = 10^{-28}$ for a future gravitational wave detector, we reiterate that, close to galactic centers in particular, the dark-matter mass density $\rho_{\rm DM}$ is expected to be larger than its local value $\rho_{\rm DM}^{\rm loc}$ by several orders of magnitude.

\section{Stellar background models}
\label{sec:background}

\subsection{Oppenheimer-Volkoff solutions}

We construct relativistic stellar models by solving the Oppenheimer-Volkoff equations (see \cite{OppV39}).  Specifically, we start with the spacetime metric in terms of 
Schwarzschild coordinates,
\begin{equation} \label{TOV_metric_areal}
ds^2 = - e^{2 \Phi(R)} dt^2 + e^{2 \lambda(R)} dR^2 + R^2 d\Omega^2,
\end{equation}
where $R$ is the areal radius, and where we adopt geometrized units with $G = 1 = c$ for the remainder of this document.  Here the metric coefficient $e^{2 \lambda}$ is given by
\begin{equation} \label{lambda}
    e^{2 \lambda} = \left(1 - \frac{2 M(R)}{R} \right)^{-1},
\end{equation}
while $\Phi$ satisfies
\begin{equation} \label{TOV_1}
\frac{d \Phi}{dR} = \frac{M(R) + 4 \pi P R^3}{R^2 - 2 M(R) R}.
\end{equation}
In the above, the (enclosed) gravitational mass $M(R)$ obeys 
\begin{equation} \label{dMdR}
\frac{dM}{dR} = 4 \pi \rho R^2,
\end{equation}
$\rho$ is the total mass-energy density,  and the pressure $P$ satisfies the (relativistic) stellar structure equation
\begin{equation} \label{TOV_2}
\frac{d P}{dR} = - (\rho + P)\frac{M(R) + 4 \pi P R^3}{R^2 - 2 M(R) R}.
\end{equation}
We solve the above equations using a polytropic equation of state (EOS)
\begin{equation} \label{eos}
P = K \rho_0^\Gamma
\end{equation}
where the rest-mass density $\rho_0$ is related to $\rho$ and $P$ by
\begin{equation}
\rho = \rho_0 + \frac{1}{\Gamma - 1} P.
\end{equation}
We denote the radius of the star, where the pressure $P$ vanishes, by $R_*$, and the total mass-energy by $M_* = M(R_*)$.

\subsection{Transformation to isotropic coordinates}
\label{sec:R_to_r}

In terms of an isotropic radius $r$, the metric (\ref{TOV_metric_areal}) can be written in the form
\begin{equation} \label{TOV_metric_isotropic}
ds^2 = - \alpha(r)^2 dt^2 + A(r)^2 (dr^2 + r^2 d\Omega^2).
\end{equation}
Matching the radial and angular coefficients with (\ref{TOV_metric_areal}) we obtain
\begin{subequations}
\begin{align}
    R & = A r \\
    e^\lambda dR & = A dr,
\end{align}
\end{subequations}
which can be combined to yield the differential equation
\begin{equation} \label{transform}
    \frac{dr}{r} = \left(1 - \frac{2 M(R)}{R} \right)^{-1/2} \frac{dR}{R}.
\end{equation}
In the exterior of the star, where $M(R) = M_*$ is constant, this equation can be integrated analytically to yield 
\begin{subequations}\label{r_R_exterior}
\begin{align} 
r & = \frac{1}{2} \left( \sqrt{R^2 - 2 M_* R} + R - M_* \right), \\
R & = r \left(1 + \frac{M_*}{2 r} \right)^2.
\end{align}
\end{subequations}
In the stellar interior, however, Eq.~(\ref{transform}) has to be integrated numerically in general, starting at the center with $r=0=R$ and fixing a constant of integration by matching to the exterior solution (\ref{r_R_exterior}) at the stellar surface.  We note that the singularity at $R = 0$ on the right-hand side of (\ref{transform}) can be handled by adding and subtracting a term $dR/R$, i.e.~by writing
\begin{align}
& \left(1 - \frac{2 M(R)}{R} \right)^{-1/2} \frac{dR}{R} = \\
& ~~~~~~~~~~~~~~~~ \frac{1 - (1 - 2 M(R)/R)^{1/2}}{R (1 - 2 M(R)/R)^{1/2}} dR + \frac{dR}{R}, \nonumber
\end{align}
where the second term on the right-side can now be integrated analytically, and the first term remains finite at $R=0$ and can therefore be integrated numerically.

We complete the transformation with the identification 
of the lapse function $\alpha(r)$ according to
\begin{equation}
\alpha(r) = e^{\Phi(R(r))}.
\end{equation}

\subsection{Constant-density star}

The limit of incompressibility ($\Gamma \rightarrow \infty$) yields stars of constant density $\rho =$ const, which can be constructed analytically.  For such stars we have 
\begin{equation} \label{mass_constant_dens}
    M(R) = \frac{4 \pi}{3} \rho R^3,
\end{equation}
so that
\begin{equation}
    e^{2 \lambda} = \left(1 - \frac{8 \pi}{3} \rho R^2\right)^{-1}.
\end{equation}
Eq.~(\ref{TOV_1}) can also be integrated analytically in this case and yields
\begin{equation} \label{Phi_0_constant_dens}
    e^{\Phi} = \frac{3}{2}\left(1 - \frac{2 M_*}{R_*}\right)^{1/2} - 
        \frac{1}{2} \left( 1 - \frac{2 M_* R^2}{R_*^3} \right)^{1/2}.
\end{equation}

For a constant-density star, the transformation (\ref{transform}) now takes the form
\begin{equation} \label{transform2}
    \frac{dr}{r} =  \frac{dR}{\left(1 - \kappa^2 R^2 \right)^{1/2} R},
\end{equation}
where we have abbreviated $\kappa^2 = 8 \pi \rho / 3$.  The expression on the right-hand side can be integrated using the substitution $1 - \kappa^2 R^2 = \tanh^2 x$, so that 
\begin{equation}
dR = - \frac{\tanh x}{\cosh^2 x} \frac{dx}{\kappa^2 R}
\end{equation}
and $\kappa^2 R^2 = 1 - \tanh^2 x = \cosh^{-2} x$, which yields
\begin{align}
& \int \frac{dR}{\left(1 - \kappa^2 R^2 \right)^{1/2} R}
= \int \frac{- \tanh x dx}{\cosh^2 x \tanh x (1 - \tanh^2 x)}  \nonumber \\
& ~~~~ = - \int dx 
= - \arctanh(\sqrt{1 - \kappa^2 R^2}) + \tilde C\nonumber \\
& ~~~~ = - \frac{1}{2} \ln \left( \frac{1 + \sqrt{1 - \kappa^2 R^2}}{1 - \sqrt{1 - \kappa^2 R^2}} \right) + \tilde C,
\end{align}
where $\tilde C$ is a constant of integration.  Combining this expression with the left-hand side of (\ref{transform2}) we obtain
\begin{align} \label{r_iso_interior}
    r(R) & = C \left( \frac{1 - \sqrt{1 - \kappa^2 R^2}}{1 + \sqrt{1 + \kappa^2 R^2}} \right)^{1/2} \nonumber \\
    & = C \left( \frac{1 - \sqrt{1 - 2 M(R)/ R}}{1 + \sqrt{1 - 2 M(R)/ R}} \right)^{1/2},
\end{align}
where $C = e^{\tilde C}$.  We can now fix this constant by requiring that the isotropic radius in the interior, given by (\ref{r_iso_interior}), matches its value in the exterior, given by (\ref{r_R_exterior}), at the stellar surface $R_*$.  This yields
\begin{equation}
C = 
r_* \left( \frac{1 + \sqrt{1 - 2 M / R_*}}{1 - \sqrt{1 - 2 M / R_*}} \right)^{1/2},
\end{equation}
where $r_* = r(R_*)$ is the isotropic stellar radius.  We can also invert (\ref{r_iso_interior}) to obtain
\begin{equation}
R(r) = \frac{1}{\kappa} \left\{1 - \tanh^2\left( \ln(r/C) \right) \right\}^{1/2}. 
\end{equation}

\section{Primordial black holes as test particles}
\label{sec:eom}

\subsection{Geodesic equation}

We track the primordial black hole's (PBH's) trajectory by treating it as a freely-falling test particle in the spacetime generated by a stellar background model as discussed in Section \ref{sec:background}.  These spacetimes posses two Killing vectors --  a time-like Killing vector $\xi_{(t)} = \partial/\partial t$ and a space-like Killing vector $\xi_{(\varphi)} = \partial/\partial \phi$ -- which give rise to two constants of motion, namely the energy per unit mass
\begin{equation} \label{define_e}
e \equiv - u_a \xi^a_{(t)} = - u_t 
= - g_{tt} \frac{dt}{d\tau}
= \mbox{const},
\end{equation}
and the angular momentum per unit mass
\begin{equation}\label{define_l}
\l \equiv u_a \xi^a_{(\varphi)} = u_\varphi 
= g_{\varphi\varphi} \frac{d\varphi}{d\tau}
= \mbox{const}.
\end{equation}
Here $u^a$ is the PBH's four-velocity, and the metric components can either be evaluated in terms of the areal radius $R$ (using the metric \ref{TOV_metric_areal}) or in terms of the isotropic radius $r$ (using the metric \ref{TOV_metric_isotropic}).   

We assume that the equatorial plane of the coordinate system is aligned with the PBH's orbit, in which case $\theta = \pi/2$, $\sin \theta = 1$, and $u^\theta = 0$.  Evaluating the geodesic equation for $u^r$ in terms of the isotropic metric (\ref{TOV_metric_isotropic}) we then obtain the equations of motion
\begin{subequations} \label{eom}
\begin{align}
\frac{dr}{dt} & = \frac{1}{A^2 u^t} u_r \\
\frac{du_r}{dt} & = - \alpha u^t \partial_r \alpha 
+ \frac{u_r^2}{u^t}\frac{\partial_r A}{A^3} + 
\frac{u_\varphi^2}{u^t} \left(\frac{1}{r^3 A^2} + \frac{\partial_r A}{r^2 A^3} \right)  \label{durdt}\\
\frac{d\varphi}{dt} & = \frac{1}{r^2 A^2 u^t} u_\varphi \\
\frac{d u_\varphi}{dt} & = 0
\end{align}
(see, e.g., \cite{ShaT85,BauS10} and note that the last equation is a direct consequence of \ref{define_l}).  In (\ref{eom}), the time component of the four-velocity $u^t$ can either be found from (\ref{define_e}), knowing $e$, or from the normalization $u_a u^a = -1$, which yields
\begin{equation} \label{ut}
\alpha u^t = \left( 1 + \frac{u_r^2}{A^2} + \frac{u_{\varphi}^2}{r^2 A^2} \right)^{1/2}.
\end{equation}
\end{subequations}

\subsection{Initial conditions}

We start all orbits by placing the PBH at an areal radius $R(0) = R_{\rm frac} R_*$ and $\varphi(0) = 0$.  From $R(0)$ we compute the initial isotropic radius $r(0)$ as described in Section \ref{sec:R_to_r}.

For a given value of $r(0)$ we compute the angular momentum $\l_{\rm circ}$ corresponding to a circular orbit by setting $u^r = 0$ and $du^r/dt = 0$ in (\ref{durdt}) and using (\ref{ut}), which results in
\begin{equation}
\l_{\rm circ} = \frac{r^3 A^3 \partial_r \alpha}{A \alpha + \alpha r \partial_r A - A r\partial_r \alpha}.
\end{equation}
We then reduce the angular momentum by a fraction $\l_{\rm frac}$ in order to obtain eccentric orbits, and start the integration with $u_\varphi(0) = \l = \l_{\rm frac} \l_{\rm circ}$.

\section{Perturbations of circular orbits}
\label{sec:perturbations}

In this section we adopt the metric in the form (\ref{TOV_metric_areal}), in terms of the areal radius $R$, in which case the two constants of motion (\ref{define_e}) and (\ref{define_l}) take the form
\begin{equation}
e = - e^{2 \Phi} \frac{dt}{d\tau}
\end{equation}
and 
\begin{equation}
\l =  R^2 \sin^2 \theta \frac{d\varphi}{d\tau}.
\end{equation}
As before we will assume that the orbit resides in the equatorial plane, so that $\theta = \pi/2$ and hence $\sin \theta = 1$ in the following.

For convenience we define a new constant $E$, which reduces to the kinetic energy of a particle at infinity in the low-velocity $v \ll 1$ limit,
\begin{align} 
E & \equiv \frac{1}{2}(e^2 - 1) = \frac{1}{2}\left( g_{tt}^2 (u^t)^2 - 1\right) \\
& = \frac{1}{2} \left( g_{tt} \left(-1 - g_{RR} (u^R)^2 - g_{\varphi\varphi} (u^\varphi)^2\right) - 1 \right) \nonumber \\
& = \frac{1}{2} \left( e^{2 \Phi} + \frac{e^{2 \Phi}}{1 - 2 M(R)/R} (u^R)^2 +  \frac{e^{2 \Phi}\l^2}{R^2} - 1 \right). \nonumber
\end{align}
Here we have used the normalization $g_{ab} u^a u^b = -1$ in going from the first line to the second, and have inserted the metric components (\ref{TOV_metric_areal}) together with (\ref{lambda}) in the last step.  We now isolate the term involving $u^R$ to obtain
\begin{align} \label{uR}
\frac{1}{2} (u^R)^2 = ~&  E e^{-2 \Phi} \left(1 - \frac{2 M(R)}{R} \right) - \frac{\l^2}{2 R^2} + \frac{M(R)\, \l^2}{R^3}  \nonumber  \\ 
& - \frac{1}{2} \left(1 - e^{-2 \Phi}\right)\left(1 - \frac{2 M(R)}{R} \right),  
\end{align}
from which we can identify the potential $V(R)$ by comparing with Eq.~(2) in the main body of the paper to find
\begin{align} \label{potential_general}
V(R) = \, & E - E e^{-2 \Phi(R)}\left(1 -\frac{2M(R)}{R}\right) - \frac{M(R)\l^2}{R^3} \nonumber \\
& + \frac{1}{2} \left(1 - e^{-2 \Phi(R)}\right) \left(1 -\frac{2M(R)}{R}\right).
\end{align}
In the following subsections we evaluate $V(R)$ in the stellar exterior, interior, as well as for constant-density stars, and then compute the periastron advance from its derivatives as discussed in the main body of the paper (e.g., Eq.~6).

\subsection{Stellar exterior}

As a way to verify our approach we first consider 
nearly circular test particle motion in the stellar exterior, 
where $M(R_*) = M_*$  and
\begin{equation}
e^{2 \Phi} = 1 - \frac{2 M_*}{R}.
\end{equation}
In this case several terms in (\ref{potential_general}) cancel conveniently and we obtain the more familiar expression,
\begin{equation}
V(R) = - \frac{M_*}{R} - \frac{M_* \l^2}{R^3}.
\end{equation}
From (6) we then have, to leading order in $\l^2$,
\begin{equation}
\frac{\tau_R}{\tau_\varphi} \simeq 1 + \frac{3 \l^2}{R^2}
\simeq 1 + \frac{3 M_*}{R},
\end{equation}
where we have used $\l^2 = M_* R^2/(R - 3 M_*)\simeq M_* R$ for a nearly circular orbit at radius $R \gg 3 M_*$ in the second estimate.  As the test particle travels from periastron to the next periastron, its positional angle $\varphi$ therefore increases by
\begin{align}
\Delta \varphi = \frac{d\varphi}{d\tau} \tau_R
= 2 \pi \frac{\tau_R}{\tau_\varphi} \simeq 2 \pi \left(
1 + \frac{3 M_*}{R} \right),
\end{align}
meaning that the periastron advances by
\begin{equation}
\Delta \varphi = \frac{6 \pi M_*}{R}
\end{equation}
as expected for a nearly circular orbit at large $R \gg M_*$.

\subsection{Stellar interior}
\label{sec:interior}

In the stellar interior the analysis is more complicated because the metric coefficients $g_{tt}$ and $g_{RR}$ are no longer inverses of each other and hence no longer cancel each other.  One approach would be to insert numerical solutions $M(R)$ and $\Phi(R)$ of the OV equations (see Section~\ref{sec:background}) into (\ref{potential_general}).  Instead, we focus our analysis to a nearly homogeneous region close to the center, where we expand the density and pressure about their central values\footnote{The $R^4$ term in the expansion of $\rho(R)$ is needed in the $M(R) \l^2/R^3$ term of (\ref{uR}) only.}
\begin{subequations} \label{expansions}
\begin{align}
\rho(R) & = \rhc + \frac{1}{2} \drhot R^2 + 
\frac{1}{24} \drhof R^4 + {\mathcal O}(R^6), \label{rho_expansion} \\
P(R) & = \Pc + \frac{1}{2} \P2 R^2 + {\mathcal O}(R^4),
\end{align}
\end{subequations}
where $\rho_c^{(n)} \equiv (d^n \rho / dR^n)_{R = 0}$ and similar for $P$.  Using (\ref{rho_expansion}) in (\ref{dMdR}) we can integrate to find the enclosed mass
\begin{equation} \label{M_int}
M(R) =\frac{4 \pi \rhc}{3} R^3 + \frac{2 \pi \drhot}{5}R^5 + \frac{\pi \drhof}{42} R^7 + {\mathcal O}(R^9).
\end{equation}
Given $M(R)$, we expand (\ref{TOV_1}) about the origin and integrate to find
\begin{align} \label{Phi_int}
\Phi(R) = ~ & \Phi_0 + \frac{2 \pi}{3} \Big(\rhc + 3 \Pc \Big) R^2 
\\ & +
\frac{\pi}{90}\Big( 9 \drhot + 240 \pi \Pc + 80 \pi \rhc^2 + 45 \P2 \Big) R^4 \nonumber \\
& + {\mathcal O}(R^6) \nonumber,
\end{align}
where $\Phi_0 = \Phi(0)$ is a constant of integration that is fixed by matching to the exterior solution at the stellar surface.  We now insert $M(R)$ and $\Phi(R)$ into (\ref{potential_general}) and expand to order $R^4$ about the origin in order to identify the coefficients $p$ and $q$ in
\begin{equation}  \label{gen_potential}
V(R) = V_0 + p R^2 + q R^4 + {\mathcal O}(R^6).
\end{equation}

We next evaluate the derivatives $\drhot$, $\drhof$, and $\P2$ that we introduced in the expansions (\ref{expansions}).  In order to obtain $\P2$ we take a derivative of the stellar structure equation (\ref{TOV_2}), replace the derivative of $M(R)$ using (\ref{dMdR}), and evaluate the result at $R = 0$ to obtain
\begin{equation} \label{Ppp}
\P2 = - \frac{4 \pi}{3} \left(\rhc^2 + 4 \Pc \rhc + 3 \Pc^2 \right).
\end{equation}
For $\drhot$ we use $\rho = \rho_0 + P/(\Gamma - 1)$ together with Eq.~(\ref{eos}) to compute
\begin{equation} \label{drhodR}
\frac{d \rho}{dR} = \frac{d\rho_0}{dR} + \frac{1}{\Gamma - 1}\frac{dP}{dR}
= \left( \frac{\rho_0^{1 - \Gamma}}{K \Gamma} + \frac{1}{\Gamma - 1}\right) \frac{dP}{dR}.
\end{equation}
We now insert (\ref{TOV_2}) on the right-hand side of Eq~(\ref{drhodR}), take another derivative, and proceed as for $\P2$ to obtain
\begin{equation} \label{rhopp}
\drhot = - \frac{4 \pi}{3} \left( \frac{\rho_{0c}^{1 - \Gamma}}{K \Gamma} + \frac{1}{\Gamma - 1}\right) \left(\rhc^2 + 4 \Pc \rhc + 3 \Pc^2 \right).
\end{equation}
Finally, the term $\drhof$ only appears in terms that we have found to be small in comparison with others; for the examples considered here it is therefore sufficient to use the leading-order Newtonian terms in two further derivatives of $\drhot$ to find
\begin{equation} \label{rhopppp}
\drhof \simeq - \frac{16 \pi^2 (\Gamma - 1) \rho_{0c}^{5 - 2 \Gamma}}{3 K^2 \Gamma^2}.
\end{equation}

Inserting Eqs.~(\ref{Ppp}), (\ref{rhopp}), and (\ref{rhopppp}) into the coefficients $V_0$, $p$, and $q$ we obtain the rather unwieldy expressions
\begin{widetext}
\begin{subequations} \label{gen_potential}
\begin{align}
V_0 & = E - e^{-2 \Phi_0} E - \frac{e^{-2 \Phi_0} - 1}{2} - \frac{4 \pi \l^2}{3} \rhc \\
p & = \frac{2 \pi e^{-2 \Phi_0}}{15} \left( 15 (\rhc + \Pc) + 10 e^{2 \Phi_0} \rhc + 30 E (\rhc + \Pc) + 4 \pi e^{2 \Phi_0} \l^2 (\rhc^2 + 4 \rhc \Pc + 3 \Pc^2)\left( \frac{\rho_{0c}^{1 - \Gamma}}{K \Gamma} + \frac{1}{\Gamma - 1} \right) \right) 
\label{gen_potential_p} \\
q & = \frac{\pi}{210} \left( 28 \pi e^{-2 \Phi_0}(\rhc^2 + 4 \rhc \Pc + 3 \Pc)  \left[  \frac{4 e^{2 \Phi_0} - 5 - 10 E}{K \Gamma} \rho_{0c}^{1 - \Gamma} -  \frac{ 5(3\Gamma - 2)(1 + 2 E) - 4 e^{2 \Phi_0}}{\Gamma - 1} \right] - 5 \l^2 \drhof \right).  \label{gen_potential_q}
\end{align}
\end{subequations}
\end{widetext}
As a consistency check we can take the Newtonian limit of (\ref{gen_potential}) to find 
\begin{subequations} \label{Newt_potential}
\begin{align}
V_{0}^{\rm Newton} & = \Phi_0\\
p_{\rm Newton} & = \frac{2 \pi}{3} \rhc \\
q_{\rm Newton} & = - \frac{2 \pi^2 \rho_c^{3 - \Gamma}}{15 K \Gamma}
= - \frac{2 \pi^2 \rho_c^2}{15 a_c^2} = \frac{\pi}{10} \drhot,
\end{align}
\end{subequations}
where we have used Newtonian expressions for the speed of sound $a^2 = \Gamma K \rho^{\Gamma - 1}$ as well as the Newtonian limit of (\ref{rhopp}), and where we do not distinguished between $\rho$ and $\rho_0$.  As expected, (\ref{Newt_potential}) agrees with the expressions that we had found from Newtonian arguments in the main body of the paper.  While it is instructive to consider the Newtonian limit, we note that the relativistic terms dominate the periastron advance for the neutron star hosts that we consider here,\footnote{This can be verified by comparing the Newtonian terms with first-order post-Newtonian terms.} so that {\em a Newtonian treatment of these effects would lead to significant errors.}

We can now compute the periastron advance from Eq.~(9) in the main body of the paper,
\begin{equation} \label{pa_gen}
\Delta \varphi \simeq - \frac{\pi}{2} \frac{q R_0^2}{p},
\end{equation}
by inserting the general relativistic expressions (\ref{gen_potential_p}) and (\ref{gen_potential_q}).  Evaluating these terms requires the values $\rhc$, $\rho_{0c}$, $\Pc$, and $\Phi_0$, which depend on the stellar background model, as well as $E$ and $\l$, which we determine for any given orbit.  Results for specific stellar models and orbits are provided Table I of the main paper, where they are compared with results from integrations of the geodesic equations of Section \ref{sec:eom}.

\subsection{Constant-density star}

For a constant-density star, both the mass $M(R)$ and the metric coefficient $e^\Phi$ are known analytically, see Eqs.~(\ref{mass_constant_dens}) and (\ref{Phi_0_constant_dens}).  Accordingly, we can insert these expressions into  (\ref{potential_general}) to obtain the potential $V(R)$ for all radii, without the need of an expansion about the origin.  Abbreviating 
\begin{equation}
\eta(R) \equiv \left(1 - \frac{2 M(R)}{R} \right)^{1/2}
= \left(1 - \frac{2 M_* R^2}{R_*^3} \right)^{1/2},
\end{equation}
and $\eta_* = \eta(R_*) = \sqrt{1- 2M_*/R_*}$ the result can be written as
\begin{align}
V(R) = \, & E - \frac{4 E \eta(R)}{(\eta(R) - 3 \eta_*)^2} - \frac{M_* \l^2}{R_*^3} \nonumber \\
& + \frac{\eta(R)}{2} \left(1 - \frac{4}{(\eta(R) - \eta_*)^2}\right).
\end{align}
Remarkably, we may therefore compute the periastron advance for nearly circular orbits directly from
\begin{equation} \label{delta_phi_exact}
\Delta \varphi = 2 \pi \left( \frac{V'}{3 V' + R V''} \right)^{1/2} - \pi
\end{equation}
for all radii.  Unlike the purely Newtonian prediction, this fully relativistic calculation leads to a non-vanishing periastron advance even for constant-density stars, in agreement with our numerical solutions to the geodesic equation of Section \ref{sec:eom}.  For the orbits listed in Table I of the main paper, for example, for which $R_{\rm frac} = 0.05$, $\l_{\rm frac} = 0.99$ and $M_*/R_* = 1/6$, Eq.~(\ref{delta_phi_exact}) yields a periastron advance of $\Delta \phi = 0.001659$.

While, for a constant-density star, an expansion about $R=0$ is not necessary, such an expansion is instructive for purposes of comparison with the approach of Section \ref{sec:interior}.  In this case we obtain
\begin{subequations}
\begin{align}
V_0 & = \frac{(1 - 3 \eta_*)^2(E - 1/2)- 4 E + 2}{(1 - 3 \eta_*)^2} + \frac{M_* \l^2}{R_*^3} \\
p & = \frac{2 M_*}{R_*^4} \,
\frac{27 M_* (\eta_* - 1) - 2 R_*(7 - 6 (E - 1) \eta_*}{(1 - 3 \eta_*)^3} \\
q & = - \frac{18 M_*^2}{R_*^6} \, \frac{(1 + 2 E) \eta_*}{(1 - 3 \eta_*)^4},
\end{align}
\end{subequations}
We may now compute the periastron advance from (\ref{pa_gen}), inserting the above expressions for $p$ and $q$.  Using this approach again for the orbit of Table I in the main paper we obtain $\Delta \varphi = 0.001655$, in very good agreement with the exact value quoted above.

\end{document}